\documentstyle[12pt]{article}

\begin{document}

\baselineskip 6mm

\begin{titlepage}

\hfill\parbox{4cm}
{ KIAS-P99008 \\ hep-th/9901152 \\ January 1999}

\vspace{15mm}
\begin{center}
{\Large \bf
Eleven-dimensional massless superparticles and matrix theory
spin-orbit couplings revisited}
\end{center}

\vspace{5mm}
\begin{center} 
Seungjoon Hyun\footnote{\tt hyun@kias.re.kr},
Youngjai Kiem\footnote{\tt ykiem@kias.re.kr},
and Hyeonjoon Shin\footnote{\tt hshin@kias.re.kr}
\\[5mm]
{\it 
School of Physics, Korea Institute for Advanced Study,
Seoul 130-012, Korea
}
\end{center}
\thispagestyle{empty}

\vfill
\begin{center}
{\bf Abstract}
\end{center}
\noindent
The classical probe dynamics of the eleven-dimensional
massless superparticles in the background geometry
produced by $N$ source $M$-momenta is investigated 
in the framework of $N$-sector DLCQ supergravity.  We
expand the probe action up to the two fermion terms 
and find that the fermionic contributions are the 
spin-orbit couplings, which precisely agree with the 
matrix theory calculations.  We comment on the lack of 
non-perturbative corrections in the one-loop matrix 
quantum mechanics effective action and its compatibility 
with the supergravity analysis.   
\vspace{2cm}
\end{titlepage}

\baselineskip 7mm

The $SU(N)$ matrix quantum mechanics, also called
matrix theory, is believed to 
provide the quantum description of the eleven-dimensional 
supergravity in the $N \rightarrow \infty$ limit \cite{bfss}.  
In the framework
of the discrete light-cone quantization (DLCQ), the 
correspondence between the matrix quantum mechanics 
and the eleven-dimensional supergravity is conjectured to 
elevate to the case of the finite $N$ and finite
eleventh circle size $R$ \cite{susskind}.
Upon taking the large $N$
limit while keeping the ratio $p_- = N/R$ fixed, the same
limit as the one in the case of the infinite momentum frame
formulation of Ref.~\cite{bfss} can be attained from the DLCQ 
framework, according to the argument of Ref.~\cite{susskind}.
Within the DLCQ framework, the precise agreement
between the matrix quantum mechanics and the classical 
eleven-dimensional massless particle
probe dynamics, corresponding to two-body $M$-momentum
scatterings, was 
demonstrated in Ref.~\cite{becker}\footnote{In Ref.~\cite{add1},
an LSZ formalism for the scattering problems
in the context of the eleven-dimensional $M$ theory was 
developed.  Recently, the general two-body scattering 
perturbative dynamics
in $M$ theory was systematically analyzed in Ref.~\cite{taylor} 
up to four fermion terms.  We note that, up to four fermion
terms, the spin effects are the spin-orbit coupling type.}.  
Specifically
the bosonic one-loop $F^4$ term computed from the
matrix quantum mechanics turns out to be identical
to the classical bosonic action $v^4$ term for the massless 
particles moving along the $M$ theory circle \cite{becker}. 

The matrix quantum mechanics has sixteen supercharges.
Therefore, the existence of the one-loop bosonic $F^4$ 
term in the effective action
implies the existence of the terms with higher 
(even) fermion numbers, up to eight fermion terms 
\cite{harvey,morales}.  Within the matrix quantum 
mechanics, by perturbative analysis, two fermion
terms \cite{kraus2}, four fermion terms \cite{mcart}
and eight fermion terms \cite{barrio} were explicitly 
calculated.  The general structure and explicit form
of the fermion terms were also analyzed from the 
ten-dimensional IIA string theory 
approach \cite{harvey,kraus2,review}
and from the eleven-dimensional supergravity 
perspective \cite{taylor}.  
A key issue to consider when approaching the problem from 
the type IIA side analysis (being the small $R$ analysis) is 
whether taking the large $N$ (thereby large $R$)
limit will induce substantial 
non-perturbative corrections.  In the case
of the D2-brane scatterings described by the
(2+1)-dimensional supersymmetric Yang-Mills theory, for 
example, there are indeed big instanton corrections in the 
large $R$ limit \cite{joe,dorey,hks2,sethi}.  A surprising
answer, at least from the matrix quantum mechanics point 
of view, is that there are no non-perturbative corrections 
as shown by the non-renomalization theorem of Sethi, Stern 
and Paban \cite{sethi2}.  Thus, the fermion terms
computed from both the  IIA theory side and the 
perturbative matrix quantum mechanics side, without any 
further corrections, should match the classical, 
eleven-dimensional $M$-momentum
scattering dynamics, including all (up to eight) fermion
terms, if the matrix quantum mechanics is to describe
the complete eleven-dimensional supergravity as 
conjectured.

Given these developments, there are at least two
issues that require better understanding.  First,
we need manifestly eleven-dimensional {\em and} 
supersymmetric description of the classical $M$
momentum dynamics, which does not rely on the 
assumption of small $R$.  This does not only give us
a framework that is valid for the large $R$ limit,
but also yields a substantial technical benefit
of the straightforward implementation of the DLCQ 
prescription, which is easy to implement in the eleven
dimensional set-up \cite{hks,hyun,hk}.
For example, in the analysis of Refs.~\cite{harvey}
and \cite{morales}, it is unclear how to implement
the DLCQ prescription and how to incorporate the
large $R$, strong coupling effects for the fermion
terms.  Second, the spin effects for the probe 
$M$-momentum should be incorporated.  In the case
of Ref.~\cite{kraus2}, the rotation effect resulting from 
the spinning source D-particles, represented by the
off-diagonal elements of the background metric
in a standard fashion, was shown to account for the
two fermion spin-orbit coupling terms of the matrix
quantum mechanics, while the probe D-particle was
assumed to be spinless.  Considering the two-body
nature of the source-probe dynamics, it will not make
a lot of difference whether we assign spins to
source D-particles or to probe D-particles, especially
as far as spin-orbit couplings are concerned.   However,
the matrix theory predicts the existence 
of the eight fermion terms \cite{barrio}, and it turns
out that the eight fermion interaction is of the spin-spin 
type; we clearly need to consider
both spinning source and spinning probe D-particles
(or $M$-momenta) at the classical level.  It is 
unclear how to incorporate the effects of the
spinning probes along the line of the supergravity
side calculation of Ref.~\cite{kraus2}.   To reproduce
eight fermion terms from the eleven-dimensional
supergravity analysis, as suggested by the membrane
static potential calculations of Ref.~\cite{vijay},
we need to consider the couplings between
the background gravitino field, produced by the
spinning source $M$-momenta, and the spin fermion 
fields of the probe $M$-momenta.  Thus, the inclusion
of the spin fermions for the probe $M$-momenta is 
essential.  

In this paper, we positively solve the two issues 
raised in the above.  We consider the superspace
formulation of the spinning massless superparticle
action in the presence of the eleven-dimensional
background geometry.   By rewriting the 
superspace fields in terms of the component fields 
up to two
fermion terms (the fermion field here represents
the spin of the probe particle), we show that the 
eleven-dimensional, classical, massless superparticle
probe action expanded up to two fermion terms,
in the presence of the spinless source $M$-momenta, 
becomes the form of the matrix quantum mechanics 
effective
action up to one-loop and two fermion terms. 
We then choose an appropriate harmonic 
function, forced upon us by the DLCQ prescription,
to describe the bosonic  background geometry
produced by source $M$-momenta.  We find that
the two actions precisely agree under this choice.  
Our set-up for the 
classical supergravity calculations is eleven-dimensional
and is able to account for the probe spin  
effects. 

Our $SO(1,10)$ spinor conventions are as follows.
The eleven-dimensional $32 \times 32$ gamma matrices 
$\Gamma^r$ ($r = 0, 1, \cdots , 9, 11$)
that we use in this paper are given by
\begin{equation}
\Gamma^0  = \left( \begin{array}{cc}
                            0 &   I_{16} \\
                    -I_{16}   & 0    \end{array} \right) ~ , ~
\Gamma^i  = \left( \begin{array}{cc}
                            0 &   \gamma^i \\
                    \gamma^i   & 0    \end{array} \right) ~ , ~
\label{gcon}
\end{equation}
\[ \Gamma^{11} = \Gamma^0 \cdots \Gamma^9 = 
 \left( \begin{array}{cc}
                            I_{16} &   0  \\
                   0  &  -I_{16}    \end{array} \right) ~ , ~
\]
where $I_{16}$ is the $16 \times 16$ identity matrix
and $\gamma^{i}$ are real $16 \times 16$ $SO(9)$ gamma 
matrices ($i=1 , \cdots, 9 $).
Our signature choice for the metric is $(-+ \cdots +)$
and the gamma matrices satisfy $(\Gamma^0)^{\dagger}
= - \Gamma^0$ and $(\Gamma^i )^{\dagger} = \Gamma^i$.  
Therefore,
for the eleven-dimensional Majorana spinors
$\chi$ and $\theta$, we have $\bar{\theta} \chi = 
\bar{\chi} \theta$ and $\bar{\theta} \Gamma^{r_1 
\cdots r_k} \chi = (-1)^{k(k+1)/2} \bar{\chi}
\Gamma^{r_1 \cdots r_k} \theta$.  Here 
$\Gamma^{r_1 \cdots r_k}$ is the totally 
anti-symmetrized $k$-product with the normalization
factor $k!$.  

We start from the description of the background
geometry of the $N$-sector DLCQ supergravity.  The 
eleven-dimensional metric is given by\footnote{For the
units chosen in this paper, we follow the notation
of Ref.~\cite{becker} combined with the
choice $2 \pi \alpha^{\prime} = 1$, which 
implies $R M_p^3 = 1$.  We further set
$M_p = R = 1$ for simplicity, which implies
$T_0 = 1$ in Ref.~\cite{kraus2}.} \cite{hks} \cite{hk}
\begin{equation}
ds_{11}^2 = dx^+ dx^- + h(r) dx^{-2} + dx_1^2 + \cdots
   + dx_9^2 ~,  
\label{11met1}
\end{equation}
where 
\begin{equation}
 h(r) = \frac{15}{2} 
  \sum_{I=1}^N \frac{1}{| \vec{r} - \vec{r_I} |^7}
\label{h}
\end{equation}
is the harmonic function of the Laplacian of the
transversal nine-dimensional space-time.  Here $\vec{r}$
denotes a nine-dimensional position $\vec{r} = (x_1 , \cdots
x_9 )$.  The asymptotically light-like coordinate
$\tau = x^+ /2$ plays the role of the 
time-coordinate and $x^-$ is compactified via the 
identification $x^- \simeq x^- + 2 \pi$
\cite{hks}.  Until the stage where we 
compare our supergravity results to the matrix theory
results, we will not make any specific assumption about
the form of the function $h(r)$ shown in Eq.~(\ref{h}).
In terms of $(t, x^{11})$ coordinates defined by 
$x^{\pm} =  x^{11} \pm t$,
the metric Eq.~(\ref{11met1}) can be rewritten as
\begin{equation}
ds_{11}^2 = - f^{-1} dt^2 + f (dx^{11} - \frac{f-1}{f} dt )^2
   + dx_1^2 + \cdots + dx_9^2 ~,  
\label{11met2}
\end{equation}
where $f = 1 + h$.  For the metric of the form 
Eq.~(\ref{11met2}), the non-vanishing spin connections
are 
\begin{equation}
\omega^0_{~i} = \omega^i_{~0} = 
\omega^{11}_{~~i} = - \omega^{i}_{~11}
= - \frac{1}{2} f^{-1/2} \partial_i f ( dx^{11} - dt )
=  - \frac{1}{2} f^{-1/2} \partial_i f dx^-
\label{spincon}
\end{equation}
\[ \omega^0_{~11} = \omega^{11}_{~~0} = - \frac{1}{2} f^{-1} 
\partial_i f dx^i ~, \]
where $i=1, \dots 9$.
The background geometry described by the metric 
Eq.~(\ref{11met2}) admits sixteen Killing spinors
that satisfy the Killing spinor equation
\begin{equation}
 D_{\mu} \eta = (\partial_\mu
- \frac{1}{4} \omega_{\mu}^{rs} \Gamma_{rs} ) \eta = 0 ~,
\label{killing}
\end{equation}
consistent with the existence of the sixteen
unbroken supersymmetries.
We use $(\mu , \nu ,  \rho , \cdots )$ indices
for the curved space bosonic
coordinates, and the tangent space bosonic
indices are represented as $(r, s, t, \cdots )$. 
Since the function $f$ depends only on the transversal 
coordinates $(x_1, \cdots, x_9)$, we can set the 
covariantly constant spinor $\eta$ as $\eta = f_{\eta} 
\epsilon_{(32)}$
where $f_{\eta}$ is a function of the transversal coordinates
and $\epsilon_{(32)}$ is a constant 32-component 
Majorana spinor.
The $\mu = 0$ and $\mu = 11$ components of Eq.~(\ref{killing})
implies that
\begin{equation} 
( \Gamma_0 + \Gamma_{11} ) \epsilon_{(32)} = 0 ~ \rightarrow ~
 \Gamma_{011}  \epsilon_{(32)} = \epsilon_{(32)} ~,
\label{t1}
\end{equation}
while the $\mu = i$ component gives
\begin{equation}
(\partial_i  + \frac{1}{4} f^{-1} \partial_i f
\Gamma_{011} ) f_{\eta} \epsilon_{(32)} = 0 ~.
\label{t2}
\end{equation}
Noting that Eq.~(\ref{t1}) implies that $\Gamma_{0 11} \epsilon
= \epsilon$, we can solve Eq.~(\ref{t2}) immediately to get
\begin{equation}
 \eta = f^{-1/4} \epsilon_{(32)} ~,~ 
\epsilon_{(32)} = 
\left( \begin{array}{c}
                         \epsilon_{(16)}  \\
          \epsilon_{(16)}     \end{array} \right)
\label{kills}
\end{equation}   
where we use the
representation of the gamma matrices that we introduced 
in (\ref{gcon}), and $\epsilon_{(16)}$ is a sixteen
component $SO(9)$ spinor. 

In the
superspace formulation in terms of the superspace 
coordinates 
\[ Z^M (\lambda ) = (x^{\mu} (\lambda) ~ , 
\theta^{\alpha} (\lambda) ) \] 
as functions of the 
world-line time coordinate $\lambda$, the manifestly 
space-time supersymmetric particle
action can be written down as follows:   
\begin{equation}
S = \int d\lambda {\cal L } = -m \int 
d\lambda ( - \eta_{rs} \Pi^r \Pi^s )^{1/2}
 \equiv -m \int d \lambda ( - g (Z))^{1/2} ~.
\label{partact}
\end{equation}
The pull-back\footnote{The indices $(A, B, C, \cdots)$ 
collectively denote the bosonic and fermionic tangent space 
indices.} $\Pi^A$ of the supervielbein\footnote{We 
use $(M, N, P, \cdots)$ to collectively denote the 
bosonic and fermionic curved space indices.} 
$E_{M}^{A}$ to the particle's
world-line satisfies $\Pi^A = ( d Z^M / d \lambda ) E_M^A$. 
The metric $\eta_{rs}$ is the Lorentz invariant
constant metric and $m$ denotes the particle mass, 
for which we will take $m \rightarrow 0 $ limit.
When taking this limit, since the background fields do 
not have any explicit dependence on $x^-$, we keep
the momentum
\begin{equation}
 p_- = \frac{\delta S}{\delta \dot{x}^-}
= \frac{m}{ (-g (Z) )^{1/2} } \eta_{rs}
\Pi^r \frac{\partial \Pi^s }{\partial \dot{x}^- }
 = {\rm fixed ~} ,
\end{equation}
where the overdot represents the derivative with
respect to $\lambda$.  This implies that as 
$m \rightarrow 0$, we have to
require $g(Z) = 0$.  Just as in Ref.~\cite{becker},
we then have to compute the Routhian
\begin{equation}
{\cal L}^{\prime} = {\cal L } - p_- \dot{x}^-
 (p_- ) ~,
\end{equation}
which reduces to
\begin{equation}
 \lim_{m \rightarrow 0} {\cal L}^{\prime}
= - p_- \dot{x}^- ~, 
\label{routh}
\end{equation}
in the massless limit.  We have to solve 
$\dot{x}^-$ from the condition $g(Z) = 0$
and plug it into Eq.~(\ref{routh}) to compute
the action.  For this purpose, we set
\begin{equation}
\Pi^r = \dot{x}^- A^r_- + B^r ~,
\end{equation}
and the condition $g(Z) = 0 $ can be solved 
for $\dot{x}^-$ in terms of $A^r_-$ and
$B$ as
\begin{equation}
\dot{x}^- = \frac{\sqrt{( \eta_{rs} A_-^r B^s )^2
- ( \eta_{rs} A_-^r A_-^s ) 
   (\eta_{tu} B^t B^u ) } - (\eta_{rs} A_-^r B^s) }
{ \eta_{rs} A_-^r A_-^s }
\end{equation}

To write down an explicit action for the 
spin-orbit couplings in a given background geometry, 
we have to expand the 
action, Eq.~(\ref{routh}), to the quadratic terms in the 
eleven-dimensional Majorana 
spinor variable $\theta$, which represents the probe 
$M$-momentum spin.
We can derive the following covariant 
expression for the superfields in terms of the component 
fields up to the quadratic terms in the fermionic variable
$\theta$;
\begin{equation}
 \Pi^r = \dot{x}^{\mu} (e_{\mu}^r 
-\frac{1}{4} \bar{\theta} \Gamma^{rst} \theta
\omega_{\mu st } )
+ \bar{\theta} \Gamma^r \dot{\theta} + \cdots ~ ,
\label{a1}
\end{equation}
which is valid when the background gravitino field
and the tensor gauge fields are zero, corresponding to the 
spinless background geometry of Eq.~(\ref{11met1}). 
Here $e_{\mu}^r$'s are the bosonic
elfbeins of the background geometry. The 
derivation of Eq.~(\ref{a1}) closely parallels the 
similar calculations in the case of the supermembranes in 
Ref.~\cite{dpp}, and relies on the simplifying feature that 
the background fields are all on-shell, satisfying the 
classical equations of motion.  Unlike the case
of the ten-dimensional D-particles where the 
super-generalization is difficult due to the existence
of the RR one-form gauge field, in the eleven-dimensional
context starting from the action (\ref{partact}), 
the long range fields that mediate interactions are all 
gravitational metric fields.  
Using Eq.~(\ref{a1}), we can compute $A_-^r$
and $B^r$ as
\begin{equation}
A^r_- = e_-^r - \frac{1}{4} \bar{\theta} \Gamma^{rst} 
\theta \omega_{-st} ~,
\label{a2}
\end{equation}
\begin{equation}
B^{r} = \dot{x}^+ ( e_+^r - \frac{1}{4}
\bar{\theta} \Gamma^{rst} \theta \omega_{+st} )
+ \dot{x}^i (e_i^r - \frac{1}{4}
\bar{\theta} \Gamma^{rst} \theta \omega_{ist} )
 + \bar{\theta} \Gamma^r \dot{\theta} ~, 
\label{a3}
\end{equation}
up to the quadratic order terms in $\theta$.  
The massless superparticle action possesses local
fermionic $\kappa$-symmetry, which in our context
can be straightforwardly verified in the einbein 
formulation.  Under the background choice of
Eq.~(\ref{11met1}), we impose the following 
$\kappa$-projection condition for the
spinor $\theta$
\begin{equation}
 (\Gamma^0 + \Gamma^{11} ) \theta = 0
~ \rightarrow ~
 \Gamma^{011} \theta = \theta
~ \rightarrow ~
  \theta = 
\left( \begin{array}{c}
                         \theta_{(16)}  \\
         - \theta_{(16)}     \end{array} \right)
\label{kappa}
\end{equation}
that results from the $\kappa$-symmetry gauge fixing. 
Here, $\theta_{(16)}$ is a sixteen component $SO(9)$
spinor.  The Killing spinors $\eta$ 
corresponding to the unbroken supersymmetry generators 
satisfy $\Gamma^{011}
\eta = - \eta$ from the Killing spinor
equation (\ref{t1}).  On the other hand, the broken
supersymmetry generators satisfy the same condition
as the projection condition imposed in Eq.~(\ref{kappa}).
In terms of the $\Gamma^{\pm} = \Gamma^{11} \pm \Gamma^0$
that satisfy $(\Gamma^+ )^2 = ( \Gamma^- )^2 = 0$
and $(\Gamma^{\pm} )^{\dagger} = \Gamma^{\mp}$,
the projection condition Eq.(\ref{kappa}) can be
expressed as 
\begin{equation}
 \Gamma^+ \theta = 0 ,
\end{equation}
which implies $\theta^T \Gamma^- = 0$.  Thus, the
Dirac conjugate $\bar{\theta} \equiv i \theta^T \Gamma^0$
can be written as
\begin{equation}
\bar{\theta} = i \theta^T \frac{1}{2} ( \Gamma^+ -
\Gamma^- ) = i \theta^T \Gamma^{\tau} ~,
\label{diracc}
\end{equation}
where $\Gamma^{\tau} = \Gamma^+ / 2$.  This computation
is consistent with our previous choice of the 
$\tau = x^+ /2 $ as a time-coordinate.

We choose the static gauge $\dot{x}^{\tau} = 1$
and we set $\dot{x}^i = v^i$. 
We then use the Majorana properties of $\theta$, the 
non-vanishing components of the spin connections shown in
Eq.~(\ref{spincon}), and Eqs.~(\ref{a2}), (\ref{a3}),
(\ref{kappa}), to explicitly compute
\begin{equation}
\eta_{rs} A^r_- A^s_- =  h + O(\theta_{(16)}^4) + \cdots ~,
\label{temp1}
\end{equation}
\begin{equation}
\eta_{rs} A^r_- B^s = 1 + 2ih f^{-1/2}
\theta^T_{(16)} \dot{\theta}_{(16)}
+ i f^{-1/2} v_i \partial_j h 
\theta^T_{(16)} \gamma^{ij} \theta_{(16)} + \cdots ~,
\label{temp2}
\end{equation}
\begin{equation}
\eta_{rs} B^r B^s = v^2 + 8i f^{-1/2} \theta^T_{(16)} 
\dot{\theta}_{(16)}
 + \cdots      ~,
\label{temp3}
\end{equation}
up to the quadratic
order terms in $\theta$ in an arbitrary BPS
background geometry of the type shown in 
Eq.~(\ref{11met1}).
Furthermore, for the small velocity expansion, we assign 
an ordering 
$O(v^i) = 1$, $O(\theta ) = 1/2$ and 
$O(d / d \lambda ) =1$.  Plugging Eqs.~(\ref{temp1}),
(\ref{temp2}) and (\ref{temp3}) into the action
(\ref{routh}) yields
\begin{equation}
{\cal L}^{\prime} = 
p_- \Big[ ~ \frac{1}{2} v^2 + 4i f^{-1/2}  \theta_{(16)}^T 
\dot{\theta }_{(16)}
\label{almost}
\end{equation}
\[ + \frac{1}{8} h (v^2 )^2 - 
\frac{i}{2} v^2 f^{-1/2} v_i \partial_j h 
 \theta_{(16)}^T \gamma^{ij}  \theta_{(16)}  
+ i v^2 f^{-1/2} h  \theta_{(16)}^T \dot{\theta}_{(16)}
 + \cdots \Big] ~, \]
where we retained terms of order up to four.
Since $\theta_{(16)}$ satisfies
$\theta_{(16)}^T \theta_{(16)} = 0$, we can absorb the
factor $f^{-1/2}$ in front of the fermion kinetic term
by rescaling $\theta_{(16)}$ via 
\begin{equation}
\psi = 2 \sqrt{2} f^{-1/4} \theta_{(16)} ~.
\label{rescale}
\end{equation}
In terms of the new $SO(9)$ spinor $\psi$, the action
(\ref{almost}) becomes
\begin{equation}
{\cal L}^{\prime} = p_-  \left[ \frac{1}{2} v^2 
+ \frac{i}{2} 
\psi^T \dot{\psi}
+ \frac{1}{8} h v^4 + 
\frac{i}{16} v^2  \partial_i h 
(\psi^T \gamma^{ij} \psi ) v_j
+ \frac{i}{8} v^2  h \psi^T \dot{\psi}
 + \cdots \right] ~.
\label{final}
\end{equation}
In general, the classical BPS 
solution space described by our metric Eq.~(\ref{11met1}) 
has the form of $N$-copy symmetric product, $S^N ( R^9 )$,
corresponding to the multi-center $N$ $M$-momentum
solutions.  The terms of order $O(v^2)$ of Eq.~(\ref{final}),
however, has the standard flat space-time normalization 
regardless of the choice of the harmonic function $h$.  
From the matrix theory point of view, it was shown in 
Ref.~\cite{sethi2} that the supersymmetric Yang-Mills 
quantum mechanics moduli space is flat, even at the 
non-perturbative level.  This behavior from the matrix
theory side is thus consistent with the first two terms
of the action (\ref{final}).   
From Eq.~(5) of Ref.~\cite{kraus2}, we have
\begin{equation}
S_0 = \int dt \Big[ -1 + \frac{1}{2} \vec{v} \cdot \vec{v}
+  \frac{15}{16} \frac{(\vec{v} \cdot \vec{v} )^2}
{r^7}
- \frac{105}{16} \frac{(x^i J^{ij} v^j )(\vec{v} \cdot
\vec{v} ) }{r^9} +  \cdots \Big] ~.
\label{per}
\end{equation}
We observe that Eq.~(\ref{per}) is the same as
Eq.~(\ref{final}), once we choose $h(r) = (15/2) (1/r^7)$ 
and identify
\begin{equation}
 J^{ij} = \frac{i}{2} \psi^T \gamma^{ij} \psi ~,
\end{equation}
except for the first constant term in Eq.~(\ref{per})
and the fermion derivative terms in Eq.~(\ref{final}).
The spin-orbit term in the effective action (\ref{per}) 
was already shown in Ref.~\cite{kraus2} to agree with 
the matrix theory two fermion terms. 

In the context of the membrane dynamics in the DLCQ 
supergravity, the classical eleven-dimensional
supergravity calculations capture the non-perturbative
instanton corrections of the (2+1)-dimensional 
supersymmetric Yang-Mills theory \cite{new}.  
For particle dynamics,
however, the eleventh direction is part of the 
longitudinal space-time of the $M$-momenta metric
(\ref{11met1}) in sharp contrast to the
membrane case analyzed in Ref.~\cite{new}.  
Thus, the eleven-dimensional
supergravity calculation, Eq.~(\ref{final}), produces 
the same result as the ten-dimensional IIA D-particle
calculations, Eq.~(\ref{per}).  From the matrix theory
point of view, the type IIA calculations and the perturbative
matrix quantum mechanics 
calculations \cite{taylor}-\cite{review} are not
corrected by the non-perturbative effects \cite{sethi2}.

The background geometry favored by
the DLCQ prescription was shown to possess sixteen
Killing spinors.  From our approach, it is in principle
straightforward to further include the 
spinning source effects (thus, to compute eight fermion
terms from supergravity), even though the actual 
computations will be considerably involved.  We can 
simply
turn on the gravitino fields (by acting the broken
supersymmetry transformations (with the generator
$\chi$) on the bosonic
background fields, Eq.~(\ref{11met1})) and, thereby, the 
off-diagonal
metric elements produced by the spinning source 
$M$-momenta, proportional to $J^{ij} = \frac{i}{2} 
\chi^T \gamma^{ij} \chi$.  In other words, we expect  
that, similar to Ref.~\cite{vijay},
the detailed forms of the gravitino and metric fields
can be obtained by the supersymmetric variation 
by the sixteen broken supersymmetry generators.
If $J_{ij}$ had the purely bosonic origin, it would
be difficult to imagine that only the terms up to
$J_{ij}^n$, where $n$ is finite, show up in the 
effective action, as the matrix theory analysis shows. 
Further works along this line are in progress.

\begin{center}
{\bf Acknowledgements}
\end{center}

We would like to thank Sangmin Lee for useful discussions.

\end{document}